\documentclass{article}
\usepackage{spconf,amsmath,graphicx}

\usepackage{amsfonts}
\usepackage{multirow}
\usepackage{hyperref}
\usepackage{subfigure}
\usepackage{booktabs,setspace}
\usepackage{threeparttable}
\usepackage{mathrsfs}
\usepackage{algorithm}
\usepackage{algpseudocode}
\ninept
\usepackage{setspace}

\usepackage{amssymb,CJK, indentfirst,balance,appendix}

\title{THE VOLCSPEECH SYSTEM FOR THE ICASSP 2022 MULTI-CHANNEL MULTI-PARTY MEETING TRANSCRIPTION CHALLENGE}
\name{\begin{tabular}{c}Chen Shen, Yi Liu, Wenzhi Fan, Bin Wang, Shixue Wen, Yao Tian, Jun Zhang, \\
Jingsheng Yang, Zejun Ma\end{tabular}}
\address{Bytedance AI Lab}

\begin{document}

%
\maketitle

\begin{abstract}
This paper describes our submission to ICASSP 2022 Multi-channel Multi-party Meeting Transcription (M2MeT) Challenge. For Track 1, we propose several approaches to empower the clustering-based speaker diarization system to handle overlapped speech. Front-end dereverberation and the direction-of-arrival (DOA) estimation are used to improve the accuracy of speaker diarization. Multi-channel combination and overlap detection are applied to reduce the missed speaker error. A modified DOVER-Lap is also proposed to fuse the results of different systems. We achieve the final DER of 5.79\% on the Eval set and 7.23\% on the Test set. For Track 2, we develop our system using the Conformer model in a joint CTC-attention architecture. Serialized output training is adopted to multi-speaker overlapped speech recognition. We propose a neural front-end module to model multi-channel audio and train the model end-to-end. Various data augmentation methods are utilized to mitigate over-fitting in the multi-channel multi-speaker E2E system. Transformer language model fusion is developed to achieve better performance. The final CER is 19.2\% on the Eval set and 20.8\% on the Test set.
\end{abstract}


\section{Introduction}

Recently, multi-channel multi-party meeting transcription has attracted increasing research interest. The speech-processing system is required to handle the complex acoustic conditions in the meeting scenario. In this paper, we introduce our speaker diarization and automatic speech recognition (ASR) systems designed for the M2MeT challenge. Considering the clustering-based speaker diarization is widely used in commercial applications, we explore different approaches to improve the performance of the clustering-based system for speech with a high speaker overlap ratio. For the ASR task, we propose several approaches to improve the accuracy of the far-field multi-speaker overlapping speech.

The organization of this paper is as follows. The details of our speaker diarization system are introduced in Section 2. Section 3 describes our end-to-end ASR system with the neural front-end and SOT. Section 4 concludes the paper.

\section{Track 1: Speaker Diarization}
\subsection{Data Preparation}

Our speaker diarization system for Track 1 consists of several blocks. The data description is as follows:

\begin{itemize}
  \item \textbf{Front-end processing}: The direction-of-arrival estimator is trained on the multi-channel data simulated by the near-field recordings of AliMeeting Corpus \cite{alimeeting} with different room impulse responses (RIRs).
  \item \textbf{Speaker embedding}: We use CN-Celeb \cite{cnceleb1, cnceleb2} as the training set. Also, the long audio in AISHELL-4 \cite{aishell4} and AliMeeting is split into short segments. Each segment only contains a single speaker and no overlap is included.
  \item \textbf{Clustering}: The PLDA is trained using CN-Celeb. The training set of AliMeeting is used as the development set to tune the parameters of VBx.
  \item \textbf{Overlap detection}: The overlap detection models are trained on the same data with the DOA estimator.
  \item \textbf{System fusion}: The parameters used in the system fusion are tuned in the Eval set of AliMeeting.
  \item \textbf{Data augmentation}: MUSAN \cite{musan} and RIRs \cite{rirs} are used in the training of the speaker embedding extractors. 
\end{itemize}

\subsection{Detailed descriptions}
\subsubsection{Front-end Processing}
Due to the high reverberation level in the distant speech communication scenarios, we use a multi-channel dereverberation algorithm based on Kalman filtering \cite{braun2021low}. The method is implemented with a STFT using 25\% overlapping 64~ms square-root Hann windows and a 1024-point FFT on 16~kHz sampled signals. The dereverberation filter length is 10 for each frequency band. 

The DOA of the sound source is proved to be helpful \cite{zheng2021real}. We train a neural-net-based DOA estimator to obtain a 36-dim probability vector representing the azimuth angles that divide the space with ten-degree intervals. We use four 2D-convolution blocks to extract frame-level features from the multi-channel input and a max-pooling layer is employed after every convolution block. Four layers of the DFSMN module \cite{bi2018deep} with a sigmoid function are implemented to produce the posteriors of the DOA.  In our experiments, the input is the 8-channel audio partitioned by 128~ms. The acoustic feature is 129-dim STFT with a frame length of 16~ms and a frame shift of 8~ms.

\subsubsection{Speaker Embedding}

Two different architectures, ResNet-101 \cite{resnet} and ECAPA-TDNN \cite{ecapa_tdnn}, are employed as the speaker embedding extractor. The detailed structure about ResNet-101 can be found in \cite{landini_vbx}. A 3-fold speed augmentation is performed so each segment is perturbed by 0.9 and 1.1 speed factors. This increases the number of training speakers to 3. The Kaldi-based offline data augmentation is then applied. We remove the babble noise since it contains human voice, which is prohibited in this challenge. To train the ECAPA-TDNN network, the SpeechBrain toolkit is used \cite{speechbrain}. Online data augmentation is implemented in this case. The babble noise is removed as well. 

For both architectures, 80-dim log Mel filter-bank energies are used as the input acoustic features. The frame size is 25~ms, and the frame shift is 10~ms.

\subsubsection{Clustering}

In the clustering stage, VBx \cite{landini_vbx} and spectral clustering are employed. For VBx, the parameters are automatically tuned on the development set using the Optuna toolkit \cite{optuna}. For spectral clustering, we use an auto-tuning version \cite{park_auto_spectral_clustering} so that the parameter is self-tuned. The clustering is performed on each audio channel.

In our experiments, we find that combining the results from all 8-channels improves the performance significantly. In the overlapped regions, different speakers might be dominant in different channels due to the directional microphone array, so the speakers could be recognized in different channels. In this way, the combination reduces the missed speaker error. The algorithm is shown in Alg. \ref{alg:multi_channel_combination}.

\begin{algorithm}
\caption{The combination of different diarization results}
\label{alg:multi_channel_combination}
\begin{algorithmic}
\Require The diarization result $\pmb{H}_c$ from channel $c=1,\ldots,8$, the empty combination $\tilde{\pmb{H}}=\phi$
\State Determine the number of speakers $N$ with majority voting.
\For {$c=1$ to 8}
\State Skip this result if the number of speakers $N_c \neq N$.
\If {$\tilde{\pmb{H}}=\phi$}
\State $\tilde{\pmb{H}}=\pmb{H}_c$
\Else
\State Find the speaker label mapping: $\pmb{H}_c \rightarrow \tilde{\pmb{H}}_c$ to minimize the diarization error rate between $\tilde{\pmb{H}}_c$ and $\tilde{\pmb{H}}$
\State $\tilde{\pmb{H}} \gets \tilde{\pmb{H}} \cup \tilde{\pmb{H}}_c$
\EndIf
\EndFor

\end{algorithmic}
\end{algorithm}

We also explore the effectiveness of the azimuth information in speaker diarization. More details will be released in the future. 

An approach is proposed to integrate the DOA with the VBx framework. The main differences of the proposed VBx are listed below.

\begin{itemize}
  \item \textbf{State-specific distribution}: We modify the state-specific distribution from $p(\pmb{x}_t|s)$ (Eq. (8) in \cite{landini_vbx}) to $p(\pmb{x}_t|s) p(\mathbf{d}_t|\mathbf{d}_s)$, where $d_t$ denotes the DOA information at time $t$ and $d_s$ represents the DOA of speaker $s$. After the initial AHC, the speaker-specific $d_s$ can be computed by averaging the DOAs belonging to the same speaker. We assume $p(\mathbf{d}_t|\mathbf{d}_s)$ follows a simple Gaussian distribution $N(\mathbf{d}_t | \mathbf{d}_s, \sigma^2 \mathbf{I})$, where $\sigma$ is a tunable parameter. We set $\sigma=0.01$ in the experiment.
  
  \item \textbf{Transition probability}: In the original VBx, the transition probability is determined by $P_{loop}$. In our implementation, we represent the DOA information by $a_t = \arg \max_i d_t(i)$, where $a_t$ denotes the most possible speech direction. When $a_{t-1} \neq a_{t}$, it indicates a speaker change. In this case, we set $p(s|s)=0.01$ and $p(s|s')=0.99, \forall s \neq s'$. Otherwise, $p(s|s)=0.99$ and $p(s|s')=0.01, \forall s\neq s'$. The probability is re-normalized per-state after the assignment.
\end{itemize}

More details and results about the DOA-enhanced VBx will be released in the future. 

\subsubsection{Overlap Detection}

We train two overlap detection models. The first model (O1) consists of a complex 2D-convolution \cite{hu2020dccrn} to handle spatial information, five separable 2D-convolution modules for feature extraction and two gated recurrent unit (GRU) layers as the back-end. The second model (O2) consists of a complex 2D-convolution, a ResNet-based front-end and two Long short-term memory (LSTM) layers as the back-end. The input is the 8-channel 129-dim STFT with a frame length of 16~ms and a frame shift of 8~ms.

The frame-level posteriors of these two overlap detectors are averaged to further improve the performance. The threshold for overlap decision is set to 0.5. A minimum silence duration of 300~ms and a minimum overlap duration of 100~ms are set to optimize the result in the development set.

\begin{table}
\centering
\begin{tabular}{lccc}\hline
Model & Precision & Recall & F1 \\ \hline
O1 & 89.4\% & 65.1\% & 75.3\% \\
O2 & 91.3\% & 60.6\% & 72.9\%\\ \hline
\end{tabular}
\caption{The results of the overlap detection on the Eval set of AliMeeting.}
\label{overlap_detection_result}
\end{table}

\subsubsection{System Fusion}
In DOVER-Lap \cite{doverlap}, the number of speakers in the overlapped region is determined by $\mathrm{round}(\sum_{s} w_s)$, where $w_s$ is the voting weight of speaker $s$. We find this operation often under-estimates the number of speakers. In our modified DOVER-Lap, a speaker is recognized in the result once the corresponding $w_s>0.5$. The voting weight is determined by a rank-based fashion and no more tuning is needed.

\subsection{Results}

\begin{table}
\centering
\begin{tabular}{lccc}\hline
  \multirow{2}{*}{VBx System} & \multicolumn{2}{c}{Eval} \\ \cline{2-3} 
  & DER(\%) & JER(\%) \\ \hline
  Baseline & 15.24 & -  \\ \hline
  VBx (ours) & 13.89 & 26.17 \\
  \ \ + multi-channel & 8.93 & 18.68  \\ 
  \ \ \ \ + DRB & 8.35 & 18.00 \\ 
  \ \ \ \ \ \ + DOA est. & 7.80 & 17.49   \\ 
  \ \ \ \ \ \ \ \ + OVD & 6.76 & 16.70  \\ 
  \ \ \ \ \ \ \ \ + OVD fusion & \textbf{6.60} & \textbf{16.70}  \\
\hline
\end{tabular}%
\caption{The DER and JER of the VBx-based system. DRB denotes the front-end dereverberation, and OVD denotes the overlap detection.}
\label{sd_single_result}
\end{table}

Table \ref{sd_single_result} shows the results of the VBx system with different configurations. The window length is 1.44~s and the window shift is 0.72~s. The DER of the official baseline is 15.24\% while our VBx baseline achieves a DER of 13.89\%. By adding different methods proposed in this paper, the DER is reduced from 13.89\% to 6.60\% on the Eval set.

\begin{table}
\centering
\resizebox{\columnwidth}{!}{%
\begin{tabular}{lccc}\hline
  \multirow{2}{*}{System} & \multicolumn{2}{c}{Eval} & Test \\ \cline{2-4} 
  & DER(\%) & JER(\%) & DER(\%)  \\ \hline
  Baseline & 15.24 & - & 15.6 \\ \hline
  1 VBx 1.44/0.72 & 6.60 & 16.70 & 8.02  \\ 
  2 VBx 1.44/0.24 & \textbf{6.44} & \textbf{16.45} & \textbf{7.74}  \\ 
  3 VBx 0.96/0.24 & 6.80 & 16.65 & 8.15  \\ 
  4 ASC 1.44/0.72 & 7.55 & 17.21 & 8.41  \\
  5 ASC 1.44/0.24 & 7.00 & 16.62 & 8.22  \\
  6 SpeechBrain retrain & 7.83 & 17.90 & 10.83  \\ \hline
  DOVER-Lap 1+2+3 & 6.29 & 16.19 & 7.57  \\
  DOVER-Lap 1+2+3+4+5+6 & 6.09 & 15.97 & 7.34  \\
  Modified DOVER-Lap & \textbf{5.79} & \textbf{15.57} & \textbf{7.23}  \\ 
\hline
\end{tabular}%
}
\caption{The results of our systems with different configurations. ASC denotes the auto-tuning spectral clustering.}
\label{sd_result}
\end{table}

Table \ref{sd_result} shows the performance of different systems based on VBx and spectral clustering. Motivated by \cite{wang2021bytedance}, systems with different time-scales are built. We first fuse the VBx-based systems. The DER is improved from 6.44\% to 6.29\%. Then, the spectral clustering is involved. The DER is further reduced to 6.09\%. With the modified DOVER-Lap, our final submission achieves a DER of 5.79\% on the Eval set and 7.23\% on the Test set.

\section{Track 2: Multi-Speaker ASR}

\subsection{Data Preparation}
We train the ASR model on AliMeeting \cite{alimeeting} and AISHELL-4 \cite{aishell4} dataset as required. A simulated dataset generated from the near-field recordings in the AliMeeting Corpus is also used for system robustness and generalization ability. The single-channel near-field clips are convolved with RIRs simulated by the image method to form the multi-channel counterpart. The T60 reverberation time ranges from 0.1~s to 1.5~s, and the room configuration is randomly generated. In order to improve the generalization of the model to overlapping speech, the number of speakers in a single clip is randomly sampled from 1 to 4, and the overlap rate distribution is similar to which of the AliMeeting corpus. The additive noise from MUSAN \cite{musan} and Freesound is also convolved with simulated RIRs, which is added to the training data at an SNR randomly sampled from 0dB to 30dB. We generate multiple copies of the data under different mixing conditions to simulate enough combinations of clean speech, background noise, and room configuration, resulting in a 600-hour simulated dataset.

We have four datasets in total for the ASR system, as shown in Table \ref{data}. D1, D2 are training sets of AliMeeting far-field and near-field audio, respectively. D3 is AISHELL-4 corpus that is recorded by a far-field microphone array. The simulated dataset D4 has near-field recordings. All datasets have a 16k sample rate.

\begin{table}
\centering
\resizebox{\columnwidth}{!}{%
\begin{tabular}{lcccc}\hline
  \multirow{2}{*}{Set} & \multicolumn{3}{c}{Training set} \\ \cline{2-4} 
  & Hours & channel\&sample rate & data \\ \hline
  D1 & 105h & 8\&16k & Train AliMeeting far \\ 
  D2 & 105h & 1\&16k & Train AliMeeting near \\ 
  D3 & 120h & 8\&16k & AISHELL-4 \\ 
  D4 & 600h & 8\&16k & 8-channel simulation \\ \hline
\hline
\end{tabular}%
}
\caption{Data description for the ASR system.}
\label{data}
\end{table}

\subsection{System Description}
Our ASR system is based on an encoder-decoder model with joint the CTC-attention structure, and serialized output training (SOT) is adopted to overlapping speech recognition. Furthermore, we use the neural front-end speech enhancement module to model 8-channel audio, which joints training with the base ASR model. Language model fusion is also used to improve performance.

\subsubsection{E2E ASR Structure}
Joint CTC-attention model \cite{kim2017joint, vaswani2017attention} is adopted to our ASR baseline model where attention decoder and CTC (connectionist temporal classification) \cite{graves2006connectionist} block receive the acoustic information from a shared encoder. The encoder is composed of several layers of Conformer \cite{gulati2020conformer} blocks, and the decoder is stacked with transformer blocks. To take the advantages of both CTC and attention mechanisms, a multi-task learning (MTL) based loss function is derived as \cite{kim2017joint}, 
\begin{equation}\label{ASR_loss}
 \mathcal{L}  = \alpha  \mathcal{L}_{CTC} + (1-\alpha)  \mathcal{L}_{Attention}
\end{equation}
The weight $\alpha$ is a tunable parameter between 0 and 1. 

During inference, the end-to-end speech recognition system combines the CTC with attention probabilities in beam search process \cite{kim2017joint, vaswani2017attention}. The CTC probabilities supply stable alignment for the attention model.

\subsubsection{Multi-speaker Speech Recognition}
Serialized output training (SOT) \cite{kanda2020serialized} is a widely used framework for multi-speaker overlapping speech recognition, which sorts multi-speaker transcriptions by start time one after another. SOT serializes multiple references into a single token sequence, which has no limitation in the maximum number of speakers and can model the dependencies among outputs for different speakers. For example, in two-speaker case, the reference label will be given as R = \{ $r_1^1, \cdots , r_{N_1}^1, \left \langle sc \right \rangle, r_1^2, r_{N_2}^2, \left \langle eos \right \rangle$ \} , which notes that $\left \langle eos \right \rangle$ is used at the end of the sequence and the special symbol $\left \langle sc \right \rangle$ represents the speaker change which is used to concatenate different utterances. This method is also called the utterance-based first-in-first-out (FIFO) method.

\subsubsection{Neural Front-end Speech Enhancement}\label{front-end}
\begin{figure}
    \centering
    \includegraphics[scale=0.2]{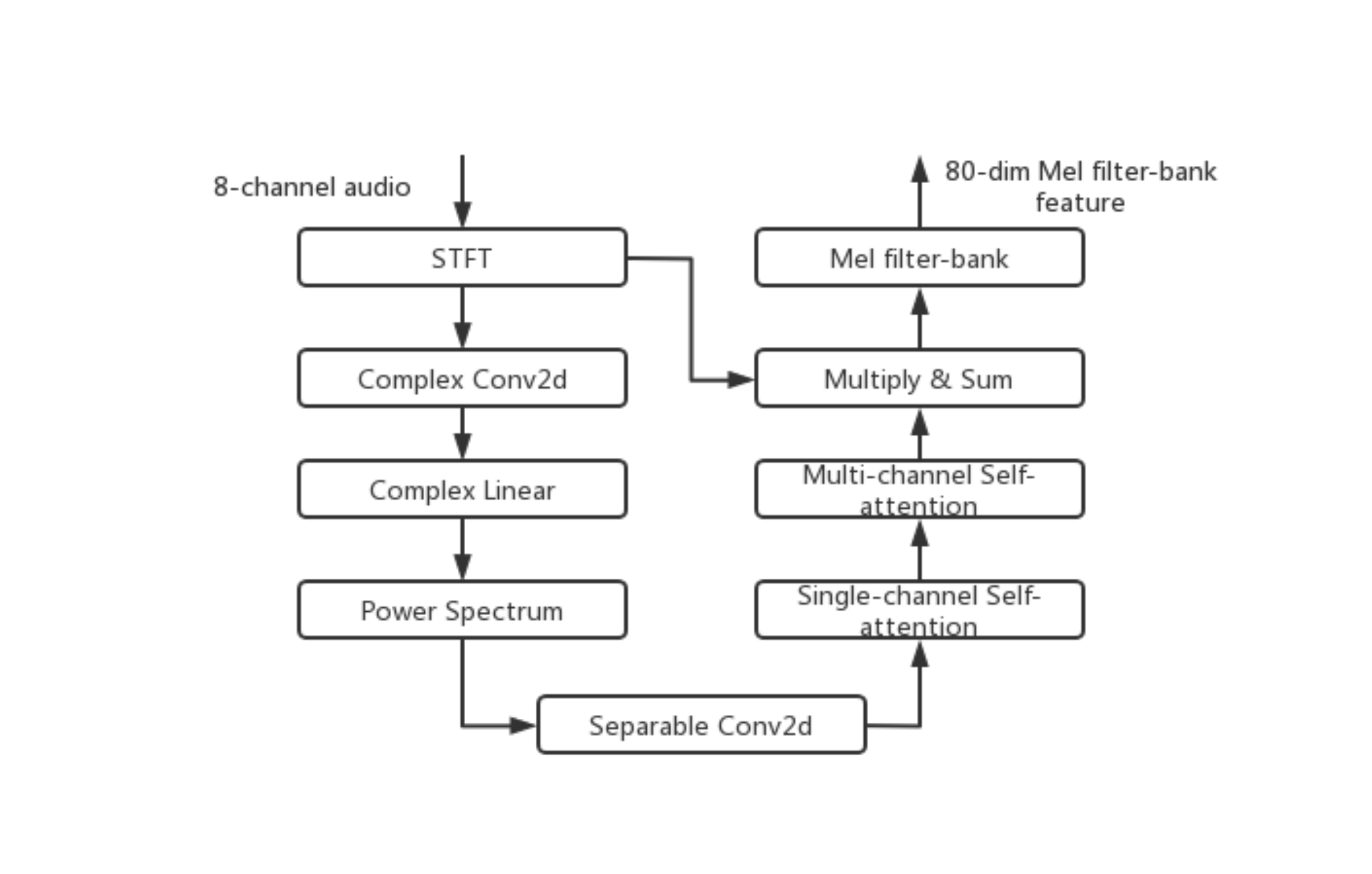}
    \caption{Neural front-end speech enhancement module}
    \label{Neural front-end speech enhancement module}
\end{figure}

It has been repeatedly reported that beamforming methods could produce substantial improvements for ASR systems \cite{wolfel2009distant} in reverberant scenarios. Recently, neural beamforming methods have drawn much attention since they have the potential to learn and adapt from massive training data, which improves their robustness to unknown positions and orientations of microphones and sources, types of acoustic sources, and room geometry. Here, an end-to-end multi-channel ASR system is adopted, bypassing above conventional beamforming formalization. The front-end module architecture is shown in Figure \ref{Neural front-end speech enhancement module}. The 8-channel audio is first input into a Complex 2D-convolution layer, and a complex linear layer \cite{hu2020dccrn} to fully exploit spectral and spatial information between different input channels. Then, the power spectrum of each channel is passed into two separable 2D-convolution blocks \cite{howard2017mobilenets}. We use one self-attention \cite{vaswani2017attention} layer for the feature of each channel. Then we concatenate the 8-channel features and use another self-attention layer with a sigmoid function to obtain a mask combining the multi-channel spectral information. Subsequently, the mask is implemented on the 8-channel STFT, and a sum operation along the channel dimension is adopted to obtain a single channel spectral feature. Finally, 80-dim Mel filter-bank energies are computed as input features for the ASR back-end.

\subsubsection{Language Model}
N-gram LM and transformer LM are adopted to our system to improve recognition performance further. Near-field transcriptions and non-overlapping transcriptions are used for language models training. Token-based units are adopted to LM modeling. During inference, language models are combined with shallow fusion. The process can formulate as
\begin{equation}
\begin{aligned}
\label{ASR_loss_with_lm}
 \hat{Y}  = \ & \text{argmax}_{Y\in y^{\ast}} \{ \alpha  P_{CTC}(Y|X) \\
 & + (1-\alpha)  P_{Attention}(Y|X) \\
 & + \beta P_{LM}(Y) \}
\end{aligned}
\end{equation}

\subsection{Results}
\subsubsection{End-to-End ASR training}
We use the ESPnet toolkit \cite{watanabe2018espnet} to build our system. We use 80-dimensional filterbank (F-bank) features as the input feature, and frame length is 25 ms, and frame shift is 10ms. The system label unit is Chinese characters and English characters, including 3694 distinct units and three special symbols that represent “unknown”, “null”, and “end of sentence” respectively.

The baseline single-channel ASR is a Conformer based AED (attention-based encoder-decoder) model, which contains a 12-layer encoder and 6-layer decoder; the self-attention and the feed-forward sub-layers have 256 and 2048 hidden units. The head number of multi-head attention is set to 4 in all attention sub-layers. The task weight $\alpha$ is empirically set to 0.3. Dropout rate is set to 0.1, and SpecAugment \cite{park2019specaugment} is applied to prevent over-fitting. We use the adam optimizer. The whole network is trained for 100 epochs and warmup is used for the first 25,000 iterations. The labels are sorted by utterance-based FIFO, and it only uses the first channel as input.

The end-to-end multi-channel model has the same backbone as the single-channel model, but with an additional front-end network which explains in section \ref{front-end}. The complex 2D-convolution has eight filters and a kernel size of [5,5]. The complex linear layer has 257 hidden units. The single-channel self-attention and multi-channel attention have 128 and 1024 hidden units, respectively. During training multi-channel model, we apply global CMVN (cepstral mean and variance normalization) on the output layer of front-end module, which is calculated by all single-channel filterbank features. 

\subsubsection{Results and analysis}
Table \ref{single_multi_cmp} shows the comparison of the single-channel model and multi-channel model. Neural front-end joint training with ASR model has 20.5\% and 20.3\% CER relative reduction for with and without simulation data respectively. We divide the evaluation set into two parts, the overlapping and the non-overlapping utterances. It is clearly observed that the neural front-end has a significant gain in overlapping speech recognition, which learns the contextual relationship within and across channels while modeling acoustic-to-text mapping.

The effect of data augmentation is also shown in Table \ref{single_multi_cmp}. We add 8-channel simulation data for training, and the single-channel model has 13.5\% CER relative reduction, and the multi-channel model has 13.3\% CER relative reduction respectively. The result indicates that matching data simulation methods benefit the model more.

\begin{table}
\centering
\resizebox{\columnwidth}{!}{%
\begin{tabular}{lcccc}\hline
  \multirow{2}{*}{ASR System} & \multicolumn{3}{c}{Eval} \\ \cline{2-4} 
  & CER(\%) & o-CER(\%) & no-CER(\%) \\ \hline
  Baseline & 29.7 & - & - \\ 
  Single-channel (D1+D2+D3) & 30.2 & 40.3 & 14.2 \\ 
  Single-channel (D1+D2+D3+D4) & 26.1 & 31.3 & 13.8 \\ \hline
  Multi-channel (D1+D3) & 24.0 & 30.7 & 13.5 \\ 
  Multi-channel (D1+D3+D4) & 20.8 & 25.9 & 12.0 \\ \hline
\hline
\end{tabular}%
}
\caption{Comparison of single-channel model and multi-channel model. o-CER means CER of overlapping utterances, and no-CER means CER of non-overlapping utterances}
\label{single_multi_cmp}
\end{table}

Finally, we use model average and LM fusion to have further improvement. As shown in Table \ref{ASR_final}, we average 20 lowest loss models, which achieves CER 6.7\% reduction relatively. We try N-gram LM and transformer LM fusion, but only transformer LM with 0.1 interpolation parameter has 1\% relative reduction. 
 
\begin{table}
\centering
\begin{tabular}{lcccc}\hline
  \multirow{2}{*}{ASR System} & \multicolumn{1}{c}{Eval} & \multicolumn{1}{c}{Test} \\ \cline{2-3} 
  & CER(\%) & CER(\%) \\ \hline
  Baseline & 29.7 & 30.9 \\ \hline
  \ \ + sim data \& sp \& rp & 24.0 & 25.1 \\ 
  \ \ \ \ + multi-channel model & 20.8 & 21.7 \\ 
  \ \ \ \ \ \ + model average & 19.4 & 20.9 \\
  \ \ \ \ \ \ \ \ + LM fusion & \textbf{19.2} & \textbf{20.8} \\
\hline
\end{tabular}%
\caption{Comparison of effective methods to improve the ASR performance. }
\label{ASR_final}
\end{table}

\section{Conclusions}

This paper describes the details of our systems built for the M2MeT challenge. For Track 1, different efforts have been made to improve the clustering-based system. Front-end dereverberation, multi-channel combination, clustering with DOA information, overlap detection and the modified DOVER-Lap are presented.
For Track 2, a joint CTC-attention Conformer-based E2E network with serialized output training is adopted to the multi-speaker ASR system. Neural front-end, data augmentation, and different LMs are investigated to achieve the best performance.


\bibliographystyle{IEEEbib}
\bibliography{strings,refs}

\end{document}